# Maximum-Entropy Revisited

**LONG V. LE,**[1†] **TAE J. KIM,**[2†] **YOUNG D. KIM,**[2,4] **D. E. ASPNES**[3,5]

[1] *Institute of Materials Science, Vietnam Academy of Science and Technology, Hanoi 100000, Vietnam*
[2] *Department of Physics, Kyung Hee University, Seoul 02447, Republic of Korea*
[3] *Department of Physics, North Carolina State University, Raleigh, NC 27695-8202 USA*
*†Authors contributed equally*
[4] *ydkim@khu.ac.kr*
[5] *aspnes@ncsu.edu*

**Abstract:** For over five decades the procedure termed maximum-entropy (M-E) has been used to "sharpen" structure in spectra, optical and otherwise. However, this is a contradiction: by modifying data, this approach violates the fundamental M-E principle, which is to extend, in a model-independent way, trends established by low-index Fourier coefficients into the white-noise region. The Burg derivation, and indirectly the prediction-error equations on which sharpening is based, both lead to the correct solution, although this has been consistently overlooked. For a single Lorentzian line these equations can be solved analytically. The resultant lineshape is an exact autoregressive model-1 (AR(1)) replica of the original, demonstrating how the M-E reconstruction extends low-index Fourier coefficients to the digital limit, and illustrating why this approach works so well for lineshapes resulting from first-order decay processes. By simultaneously retaining low-index coefficients exactly and eliminating Gibbs oscillations, M-E noise filtering is quantitatively superior to that achieved by any linear method, including the high-performance filters recently proposed. Examples are provided.

## 1. Introduction

The goal of spectroscopy is information, and enormous effort has gone into improving capabilities either by enhancing structure or reducing noise. Noise reduction, either linear or nonlinear, depends on information residing in correlations between data points whereas noise is uncorrelated. Hence information is contained in low-order Fourier coefficients, whereas high-order coefficients are dominated by noise.

Extracting information from spectra by linear methods generally depends on noise reduction, whether to obtain cleaner results for inspection, fit model lineshapes with less uncertainty, or enhance structure by differentiation, either as an intrinsic part of data acquisition or by numerical differentiation afterward. Linear methods reduce noise by attenuating or eliminating high-order coefficients either through reciprocal- (Fourier-) space (RS) filtering or direct- (spectral-) space (DS) convolution. While noise is reduced, the associated cutoffs generate Gibbs oscillations, creating new problems.

By now, linear methods are well understood. In particular, recent work has shown that the optimum linear approach for preserving information while reducing noise is done with high-performance filters such as cosine-terminated [1] or high-order Gauss-Hermite filters [1–5], with parameters established by the RS behavior of the data [1,3–5]. The best filters avoid the Butterworth approach of eliminating as many low-order derivatives as possible in RS expansions of transfer functions about $n = 0$, where $n$ is the coefficient index, in favor of engineering improved cutoffs. However, Gibbs-oscillation concerns remain.

Nonlinear approaches offer new capabilities, for example retaining low-index coefficients exactly but eliminating Gibbs oscillations by extrapolating trends established by these coefficients into informationally inaccessible regions. For example, one can least-squares fit model functions to spectra in DS or coefficients in RS, in the latter case using the fitted functions to replace coefficients in the white-noise region [5].

In principle the maximum-entropy (M-E) method offers a better alternative, because extrapolations are done in a most-probable, model-independent way. However, despite the fact that the correct approach follows directly from Burg's work [6,7] in the 1960's and indirectly from the prediction-error work of Kolmonov and Wiener [8–10] on weakly stationary time series in the 1940's, this has never been explored. From the beginning, what is termed M-E has been used to "sharpen" or "whiten" structure in spectra [11–14], thereby violating the fundamental M-E principle of keeping the data intact. While Burg's derivation is more relevant to filtering than the W-K approach of extracting a single harmonic buried in weakly stationary noise, the latter theory was already well developed when Burg did his work, so he modified the final step of his derivation to conform to the prediction-error result. By now the prediction-error method has been thoroughly investigated empirically [7,11,15,16] and used in many fields, including for example geophysics [12], astronomy [17,18], X-ray photoelectron spectroscopy [13,19–23], Raman scattering [14,24–28], and others [29–31]. We have found only one reference [14] where the correct approach has even been mentioned, let alone used.

Our goals in the present work are to investigate the properties of the overlooked solution, understand the mechanism by which it functions, and to use it to reduce or even eliminate noise in spectra. For a single Lorentzian structure we find that the M-E equations can be solved exactly. This solution involves only the two lowest-index coefficients, and yields an exact autoregressive replica (AR(1) in prediction-error terms) of the original. Replication is the result of a combination of two nonlinearities to form a third, the approximate Lorentzian itself. This solution explains not only why the correct M-E procedure reconstructs digital Fourier coefficients to all accessible orders, but also why it applies so well to spectra with structure arising from first-order decay processes. By retaining low-order coefficients exactly and eliminating Gibbs oscillations, the correct M-E solution is quantitatively superior to any form of linear filtering, with minimal additional computational effort.

## 2. Theory

### 2.1 Fourier analysis

The M-E approach is based on digital Fourier analysis. Because digital methods are intrinsically periodic, data should be pre-processed with either internal [32] or external [33] methods of removing principal components resulting from endpoint discontinuities in either value or slope, so the mathematics does not treat these singularities as legitimate critical points. Because their broadening is zero, these discontinuities can exert an outsized influence on results. Second and third constraints follow in that spectra to be processed must be real and positive definite. For spectra with negative excursions this is easily fixed by adding a constant background.

To define procedures, let $P(\theta) > 0$ be a continuous function with a Fourier representation

$$P(\theta) = \sum_{n=-\infty}^{\infty} R_n e^{in\theta} . \tag{1}$$

For spectroscopic applications we suppose that the data $P_j = \sum_{n=-\infty}^{\infty} R_n e^{in\theta_j}$ are given as a finite set of $(2N+1)$ real positive values $\{P_j\}$, with $j = -N, -N+1, ..., N-1, N$, and

$$\theta_j = \frac{2\pi}{2N+1} j . \tag{2}$$

The points are assumed to be equally spaced in $x$, where $x$ can be energy, wavelength, frequency, the index $j$ as shown here, or any other variable. The ends $\pm x$ of the range occur at the nonintegral values $j = \pm(N+1/2)$, respectively, so all data are interior to the range, an advantage. The restriction to an odd number of points is not essential, but is done to simplify the mathematics. Because the $P_j = P(\theta_j)$ are real, the $R_n$ satisfy the reality condition

$$R_n = R_{-n}^*, \tag{3}$$

so in practice we need coefficients $R_n$ only for $n \geq 0$. The $R_n$ are related to the standard cosine and sine coefficients $A_n$ and $B_n$ by

$$R_0 = A_0; \quad R_{n \neq 0} = A_n - iB_n. \tag{4}$$

## 2.2 The Burg derivation

We now return to $P(\theta)$. The continuous function $P(\theta)$ cannot be constructed from data because the data provide only $(2N+1)$ coefficients, or $(N+1)$ independent values when the reality condition is applied. Because data are typically noisy, this number is optimistic, possibly highly optimistic. Accordingly, let the number of useful coefficients be $M$, where $M < N$. Our task is to find a recipe by which the inaccessible coefficients can be estimated and the continuous function $P(\theta)$ created. If we can construct a best-estimate of $P(\theta)$ based on known coefficients, then in principle a Fourier transform of $P(\theta)$ generates the desired projection of the $R_n$ into the white-noise range.

This is done as follows. Given that the information content of a bitstream is proportional to the natural logarithm of the number of bit locations in the stream, we can retain intrinsic information and prevent new information from entering, for example setting $R_n = 0$ for $n > M$, by maximizing

$$S = \int_{-\pi}^{\pi} d\theta \ln(P(\theta)) = \int_{-\pi}^{\pi} d\theta \ln\left(\sum_{n=-\infty}^{\infty} R_n e^{in\theta}\right). \tag{5}$$

Because the $R_n$ for $|n| \leq M$ are already known, the only option lies with the $R_n$ for $n > M$. Accordingly, set

$$\frac{\partial S}{\partial R_{n'}} = \frac{\partial}{\partial R_{n'}}\left(\int_{-\pi}^{\pi} d\theta \ln\left(\sum_{n=-\infty}^{\infty} R_n e^{in\theta}\right)\right) = \int_{-\pi}^{\pi} d\theta \frac{e^{in'\theta}}{\sum_{n=-\infty}^{\infty} R_n e^{in\theta}} = \int_{-\pi}^{\pi} d\theta \frac{e^{in'\theta}}{P(\theta)} = 0. \tag{6}$$

for $|n'| > M$. As Burg notes, Eq. (6) can also be obtained using Lagrangian multipliers.

Now because $P(\theta)$ is positive definite, it has no zeroes over the range $-\pi \leq \theta \leq \pi$. Thus $1/P(\theta)$ is analytic over this range, and can itself be represented as a Fourier series. Let this series be

$$\frac{1}{P(\theta)} = \sum_{n'=-\infty}^{\infty} \lambda_{n'} e^{in'\theta}, \tag{7}$$

where $\lambda_{n'} = \lambda_{-n'}^*$. Substituting Eq. (7) into the maximization condition yields

$$0 = \int_{-\pi}^{\pi} d\theta \frac{e^{in\theta}}{P(\theta)} = \int_{-\pi}^{\pi} d\theta \left(\sum_{n'=-\infty}^{\infty} \lambda_{n'} e^{in'\theta}\right) e^{in\theta} \tag{8}$$

for $|n| > M$. But this says that $\lambda_n = 0$ for $|n| \geq M$. Thus

$$\frac{1}{P(\theta)} = \sum_{n=-M}^{M} \lambda_n e^{in\theta}. \tag{9}$$

The original *infinite* series has therefore become the reciprocal of a *finite* series. The $\lambda_n$ are defined by the requirement that the Fourier coefficients $R_n$ of $P(\theta)$ for $|n| \leq M$ must equal the known values obtained from the spectrum.

Calculation of the $\lambda_n$ can be bypassed by defining new coefficients $a_n$ according to

$$P(\theta) = \sum_{n''=-\infty}^{\infty} R_{n''} e^{in''\theta} = \frac{1}{\left(\sum_{n=0}^{M}(a_n e^{in\theta})\right)\left(\sum_{n'=0}^{M}(a_{n'}^* e^{-in'\theta})\right)}, \quad (10)$$

where $a_n = a_{-n}^*$, noting that $P(\theta)$ is real. Executing the inverse transform leads to

$$\int_{-\pi}^{\pi} d\theta P(\theta) e^{-im\theta} = \int_{-\pi}^{\pi} d\theta \left(\sum_{n'=-\infty}^{\infty} R_{n'} e^{i(n'-m)\theta}\right) = 2\pi R_m \quad (11a)$$

$$= \int_{-\pi}^{\pi} d\theta \left(\frac{e^{-im\theta}}{\left(\sum_{n=0}^{M}(a_n e^{in\theta})\right)\left(\sum_{n'=0}^{M}(a_{n'}^* e^{-in'\theta})\right)}\right). \quad (11b)$$

If the index $m$ is shifted to $(m-n')$, then

$$2\pi R_{m-n'} = \int_{-\pi}^{\pi} d\theta \left(\frac{e^{-i(m-n')\theta}}{\left(\sum_{n=0}^{M}(a_n e^{in\theta})\right)\left(\sum_{n'=0}^{M}(a_{n'}^* e^{-in'\theta})\right)}\right). \quad (12)$$

Next, multiply both sides by $a_m^*$ and sum over $m$, obtaining

$$2\pi \sum_{m=0}^{M} a_m^* R_{m-n'} = \int_{-\pi}^{\pi} d\theta \left(\frac{e^{in'\theta}\left(\sum_{m=0}^{M} a_m^* e^{-im\theta}\right)}{\left(\sum_{n=0}^{M}(a_n e^{in\theta})\right)\left(\sum_{n'=0}^{M}(a_{n'}^* e^{-in'\theta})\right)}\right). \quad (13)$$

But the sum in the numerator is the same as the sum on the right in the denominator. Hence these two sums cancel, leaving

$$2\pi \sum_{m=0}^{M} a_m^* R_{m-n'} = \int_{-\pi}^{\pi} d\theta \left(\frac{e^{in'\theta}}{\sum_{n=0}^{M} a_n e^{in\theta}}\right) = \frac{1}{a_0} \int_{-\pi}^{\pi} d\theta \left(\frac{e^{in'\theta}}{1+\sum_{n=1}^{M}(a_n/a_0) e^{in\theta}}\right). \quad (14)$$

Because $P(\theta)$ is minimum-phase (has no zeroes in its range of definition), the large term in brackets can be expanded, writing

$$\left(1+\sum_{n=1}^{M}(a_n/a_0) e^{in\theta}\right)^{-1} = 1 - \sum_{n=1}^{M}(a_n/a_0) e^{in\theta} + \left(\sum_{n=1}^{M}(a_n/a_0) e^{in\theta}\right)^2 - \ldots \quad (15)$$

Now since $n \geq 1$, all terms in the above integral vanish upon integration except the first. The result is

$$\sum_{n=0}^{M} a_n^* R_{n-n'} = \begin{cases} \dfrac{1}{a_0} & \text{for } n' = 0; \\ 0 & \text{for } n' = 1, 2, \ldots, M, \end{cases} \quad (16)$$

which in matrix form is

$$\begin{pmatrix} R_0 & R_1 & ... & R_M \\ R_{-1} & R_0 & ... & R_{M-1} \\ ... & ... & ... & ... \\ R_{-M} & R_{-M+1} & ... & R_0 \end{pmatrix} \begin{pmatrix} a_0^* \\ a_1^* \\ ... \\ a_M^* \end{pmatrix} = \begin{pmatrix} R_0 & R_1 & ... & R_M \\ R_1^* & R_0 & ... & R_{M-1} \\ ... & ... & ... & ... \\ R_M^* & R_{M-1}^* & ... & R_0 \end{pmatrix} \begin{pmatrix} a_0^* \\ a_1^* \\ ... \\ a_M^* \end{pmatrix} = \begin{pmatrix} 1/a_0 \\ 0 \\ ... \\ 0 \end{pmatrix}. \qquad (17)$$

Consequently,

$$a_0 = \sqrt{(\underline{R}^{-1})_{00}} \; ; \qquad (18a)$$

$$a_\nu = \frac{1}{a_0} (\underline{R}^{-1})_{0\nu}, \quad \nu = 1, 2, ..., m; \qquad (18b)$$

noting that $a_0$ is real. Therefore

$$P(\theta) = \frac{1}{\left| \sum_{n=0}^{M} (a_n e^{in\theta}) \right|^2}, \qquad (19)$$

Burg's result differs in the following ways. At Eq. (17), Burg sets $a_0 = 1$, identifies the term $1/a_0$ in the vector on the right with the expectation value $P_M$ of the noise power of the prediction-error equivalent, and discards the last-line equation, replacing it with the requirement that in a recursive solution of Eq. (17), $P_M$ must be minimized at each step. These changes map Eq. (17) onto the prediction-error approach, except that in prediction-error calculations, the $R_n$ are autocorrelations, not Fourier coefficients. These are natural interpretations of the prediction-error formulation, which is intended to recover signal harmonics buried in noise in time-series data. The result is a significant enhancement of signal-harmonic spectra, to the extent that these can even appear as delta functions. This variant will be discussed in more detail elsewhere.

### 3. An exact solution

If $R_n = e^{-|n|\Gamma}$, that is, the continuum Fourier transform of the Lorentzian

$$F(\theta) = \frac{1}{\theta^2 + \Gamma^2}, \qquad (20)$$

we now show that Eqs. (10) and (17) have an exact solution. This leads not only to a better understanding of the M-E mechanism, but also of more general results. For example, let $M = 4$. Then Eq. (17) becomes

$$\begin{pmatrix} 1 & u & u^2 & u^3 & u^4 \\ u & 1 & u & u^2 & u^3 \\ u^2 & u & 1 & u & u^2 \\ u^3 & u^2 & u & 1 & u \\ u^4 & u^3 & u^2 & u & 1 \end{pmatrix} \begin{pmatrix} a_0 \\ a_1 \\ a_2 \\ a_3 \\ a_4 \end{pmatrix} = \begin{pmatrix} 1/a_0 \\ 0 \\ 0 \\ 0 \\ 0 \end{pmatrix}, \qquad (21)$$

where $u = e^{-|n|\Gamma}$. The individual equations are

$$a_0 + ua_1 + u^2 a_2 + u^3 a_3 + u^4 a_4 = 1/a_0 \; ; \qquad (22a)$$

$$ua_0 + a_1 + ua_2 + u^2 a_3 + u^3 a_4 = 0 \; ; \qquad (22b)$$

$$u^2 a_0 + ua_1 + a_2 + ua_3 + u^2 a_4 = 0 \; ; \qquad (22c)$$

$$u^3 a_0 + u^2 a_1 + u a_2 + a_3 + u a_4 = 0 ; \tag{22d}$$

$$u^4 a_0 + u^3 a_1 + u^2 a_2 + u a_3 + a_4 = 0 . \tag{22e}$$

These are standard difference equations, suggesting a trial solution $a_n = \alpha s^{-n}$, where $\alpha$ is a constant. Making this substitution

$$a_0 = \frac{1}{\sqrt{1-u^2}} ; \tag{23a}$$

$$a_1 = -u a_0 = -\frac{u}{\sqrt{1-u^2}} ; \tag{23b}$$

$$a_n = 0 \text{ for } |n| = 2, 3, 4 . \tag{23c}$$

The result $a_n = 0$ for $|n| > 1$ is easily shown to be general. Substituting $a_0$ and $a_1$ into Eq. (19), the result for any $M > 1$ is

$$P(\theta) = \frac{1}{|a_0 + a_1 u|^2} = \frac{1-u^2}{1+u^2 - 2u \cos\theta} \tag{24a}$$

$$= \frac{1-e^{-2\Gamma}}{1+e^{-2\Gamma} - 2e^{-\Gamma} \cos\theta} . \tag{24b}$$

Here, $P(\theta)$ depends entirely on the single parameter $\Gamma$. In prediction-error theory, Eq. (24b) is the lowest-order filter response AR(1).

The connection to Eq. (20), the original Lorentzian, can now be established. Let $\Gamma \to 0$ and consider $\theta \sim 0$. Then using the expansions $e^{-\Gamma} \cong 1 - \Gamma + \Gamma^2/2 - ...$ and $\cos\theta \cong 1 - \theta^2/2 + ...$ we find

$$P(\theta) \cong \frac{2\Gamma}{\theta^2 + \Gamma^2} . \tag{25}$$

Thus the original Lorentzian line is recovered to within a scaling factor. The nonlinearity introduced by taking the reciprocal compensates the nonlinearity inherent in $\cos\theta$ to yield a third nonlinearity, the original Lorentzian. The Fourier transform of Eq. (25) therefore projects the trend established by – in this case two – low-order Fourier coefficients into the white-noise region, which is why the correct M-E theory works. In effect, the combination of the Toeplitz matrix and Eqs. (17) and (19) do a Fourier inversion, yielding a periodic replica of the original spectrum.

### 4. Discussion

While Eqs. (24) reduce to a Lorentzian lineshape in the limit of zero broadening, there is a fundamental difference between Eq. (20) and Eqs. (10) and (24b). The latter two are periodic whereas Eq. (20) is not. Hence Eq. (24b) may be termed more accurately a pseudo-Lorentzian. One consequence of periodicity is that as $n \to N$, the $R_n$ approach a constant value. For the symmetric line shown, Eq. (24b) also approaches a constant value as $\theta$ approaches $\pm\pi$. Therefore, the Eq. (24b) replica cannot follow the Lorentzian form exactly, but deviates at the edges of the range.

Insight into the Gibbs oscillations arising in linear filtering can be obtained from the discrete Fourier transform of $R_n = e^{-|n|\Gamma}$. For $(2N+1)$ points this is

$$P_N(\theta) = 2\operatorname{Re}\left(\frac{1-e^{-(N+1)(\Gamma+i\theta)}}{1-e^{-(\Gamma+i\theta)}}\right) - 1 \tag{26a}$$

$$= \frac{1 - e^{-2\Gamma} - 2e^{-(N+1)\Gamma}\left(\cos(N+1)\theta + e^{-\Gamma}\cos(N\theta)\right)}{1 + e^{-2\Gamma} - 2e^{-\Gamma}\cos\theta}. \tag{26b}$$

Equation (26b) is identical to Eq. (24b) except for the correction term due to the truncation of the sum. The extra term describes the Gibbs oscillations that result from the "ideal" linear filter, where all RS coefficients up to a white-noise cutoff are multiplied by 1, and everything beyond by zero. While high-performance linear filters use a graded cutoff, the resulting Gibbs oscillations remain of the order of those described in Eq. (26b).

We now consider poles. Letting $e^{i\theta} = z$, in the usual $z$-transform notation, the polynomial in the denominator of Eq. (19) can be factored into a product of $M$ poles. Let one such pole $(c_0 - c_1 e^{i\theta}) \to (c_0 - c_1 z)$. Then

$$P(z) = \frac{1}{|c_0 - c_1 z|^2 \, |f_{M-1}(z)|^2}. \tag{27}$$

where $f_{M-1}(z)$ is the residual polynomial of order $(M-1)$. Therefore, $P(\theta)$ has a double pole at $z = c_0/c_1$. From Eqs. (24) $|c_0/c_1| = e^{\Gamma} > 1$. Because the calculation involves no specific pole, it applies to all. Hence by causality $P(\theta)$ is minimum-phase, with all poles lying outside the unit circle, consistent with the derivation. Moreover, $|z|$ reaches its minimum value at $\theta_z = -(\theta_1 - \theta_0)$, with $P(\theta)$ decreasing quadratically in $(\theta - \theta_s)$ away from the peak. Because $(\theta_1 - \theta_0)$ is not specified, it can be selected to locate the pole at any point in the range. It follows a minimum of $M$ poles is necessary to describe a system with $M$ features. The incorporation of extra poles acts to improve the fit to the data.

Given the above, the prediction-error approach can also be understood qualitatively. As noted above, in solving Eq. (17) identifications and procedures are changed. In particular, replacing the final equation in the matrix with the condition that the power be minimized at every recursion places the next root $z_j$ as close to the unit circle as possible, resulting in a structure that is sharper than what would be obtained from the direct solution of Eq. (17). Sharpening is thus understood.

## 5. Applications

The primary objective of the present work is noise reduction. To illustrate capabilities, we consider two sets of data: XPS results obtained on a sparse distribution of AuPt clusters on an indium tin oxide substrate [34], shown in Fig. 1, and spectroscopic ellipsometric (SE) data of a MoSe$_2$ monolayer at 31 K [35], shown in Fig. 2. The energy range of the XPS data includes the 4$f$ transitions of Pt, and are relatively noisy because only a small amount of material is available. The SE data exhibit relatively little noise, but must be differentiated twice with respect to energy to assess the possible presence of a weak structure.

We consider the XPS data first. After using the external removal of endpoint discontinuities (ERED) [30] procedure to remove endpoint-discontinuity artifacts, their Fourier transform was calculated. The result is shown in Fig. 3. This behavior is characteristic of noisy spectra containing multiple structures. The low-order coefficients are relatively noise-free and exhibit interference from the different peaks. The coefficients saturate above the white-noise cutoff, here $n_c \approx 35$. The second curve shows the M-E extrapolation for $M = 30$. The low-order coefficients are followed exactly, and the trend that is established by these coefficients is followed to the limit imposed by the accuracy of the calculations. In Fig. 1 the reconstructed spectrum is superposed on the original and is essentially noise-free.

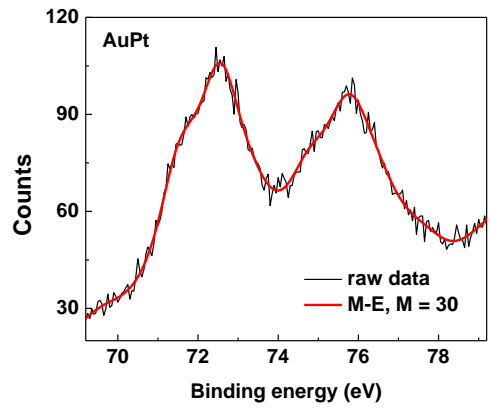

Fig. 1. Irregular curve: XPS data of AuPt clusters in the range of the Pt 4f transitions, along with its M-E reconstruction.

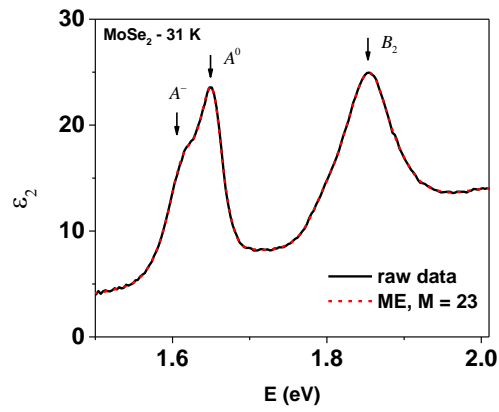

Fig. 2. Solid curve: SE data of monolayer MoSe$_2$ at 31 K, along with its M-E reconstruction.

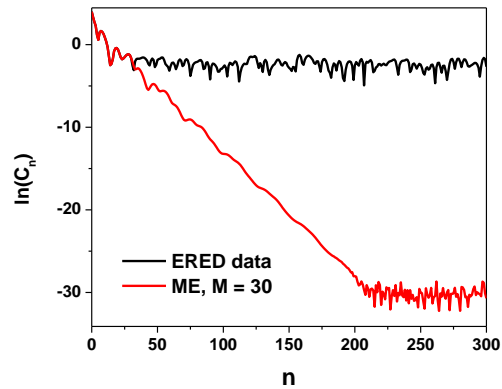

Fig. 3. Upper line: $\ln C_n$ of the XPS spectrum of Fig. 1, along with $\ln C_n$ of its M-E reconstruction for M = 30.

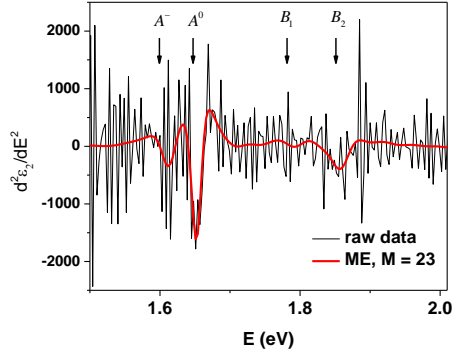

Fig. 4. Noisy spectrum: numerically calculated second derivative with respect to energy of the data of Fig. 2, along with its equivalent calculated from the M-E reconstruction. The reconstruction reveals the trion-$B_1$ structure.

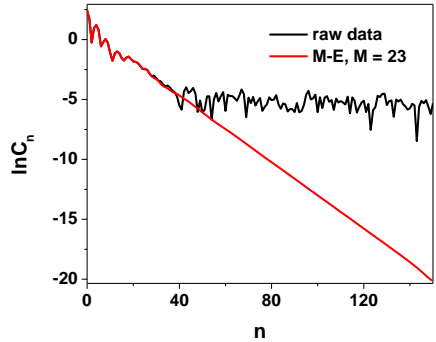

Fig. 5. $\ln(C_n)$ of the data of Fig. 2, along with $\ln(C_n)$ of its M-E reconstruction for $M = 23$.

In the MoSe$_2$ case, the data are relatively clean, but as shown in Fig. 4, the second derivative exhibits too much noise to allow meaningful conclusions to be drawn. Figure 5 shows the associated Fourier coefficients, along with the extrapolation calculated assuming a cutoff index $M = n_c = 23$. The second derivative of the reconstructed lineshape exhibits essentially no noise, revealing the $B_1$ structure

Being generated from analytic reconstructed spectra, in principle these Fourier coefficients can be extended as far as desired. This opens the possibility of self-deconvolution without apodization. Self-deconvolution was first studied by Kauppinen et al. [36] with the objective of minimizing apodization effects. We have found an analogous approach to be useful in analyzing spectra with Gaussian features, whose parabolic decrease with coefficient index $n$ is inconsistent with the above filtering procedures. By multiplying the coefficients by $e^{\alpha n^2}$, where $\alpha$ is a constant, the curvature in $\ln(C_n)$ can be removed, allowing processing to proceed. Once noise has been removed to the extent possible, the curvature can be restored and the filtered lineshape recovered. These and other topics will be discussed in a subsequent publication.

## 6. Conclusions

While the Burg and prediction -error formulations can both lead to the correct M-E solution, this aspect has been consistently overlooked. In the present work we obtain this solution, discuss its main characteristics, and apply it to different situations to demonstrate its effectiveness in removing noise. The solution for the single Lorentzian is exact, offering insight into the basic M-E mechanism, including how filtering is accomplished and how filtered lineshapes are reconstructed.

Returning to our main purpose, the model-independent M-E method of projecting trends established in low-order coefficients into the white-noise region eliminates Gibbs oscillations, the major remaining problem with high-performance linear filters. By retaining the low-index coefficients exactly, the M-E process equals the high-performance filters in this regard. The results are therefore quantitatively superior to those that can be obtained with any linear filter, with residual noise contributions at an absolute minimum.

**Funding.** National Research Foundation of Korea (NRF-2020R1A2C1009041).

**Disclosures**. The authors declare no conflicts of interest.